\documentclass[aip,jcp,reprint]{revtex4-2}

\usepackage{hyperref}
\usepackage{xurl}
\hypersetup{breaklinks=true}

\usepackage{natbib}
\usepackage[version=3]{mhchem} %
\usepackage{bm} %
\usepackage{threeparttable} %
\usepackage{dcolumn}
\usepackage{booktabs} %
\usepackage{xspace}
\usepackage{braket}
\usepackage{comment}
\usepackage{color, colortbl}
\usepackage{soul}
\usepackage{titlesec}
\titlelabel{\thetitle.\quad}
\usepackage{extarrows}
\usepackage{mathtools}
\usepackage{multirow}

\usepackage{hyperref} %
\hypersetup{colorlinks=true, citecolor=blue, urlcolor=blue, linkcolor=blue}
\usepackage{cleveref}
  	\crefname{figure}{Figure}{Figures}
  	\crefname{table}{Table}{Tables}
  	\crefname{equation}{Eq.}{Eqs.}
  	\crefname{section}{Section}{Sections}
  	\crefname{subsection}{Section}{Sections}
  	\crefname{subsubsection}{Section}{Sections}
  	\crefname{algorithm}{Algorithm}{Algorithms}

\usepackage{todonotes}

\newcommand{\code}[1]{\texttt{#1}}

\begin{document}

\title{Comment on ``Canonical transcorrelated theory with projected Slater-type geminals''  [J. Chem. Phys. 136, 084107 (2012)]} %

\author{Conner Masteran}
\affiliation{Department of Chemistry, Virginia Tech, Blacksburg, VA 24061, USA}

\author{Ashutosh Kumar}
\affiliation{Theoretical Division, Los Alamos National Laboratory, Los Alamos, New Mexico 87545, United States}

\author{Nakul Teke}
\affiliation{Department of Chemistry, Virginia Tech, Blacksburg, VA 24061, USA}

\author{Bimal Gaudel}
\affiliation{Department of Chemistry, Virginia Tech, Blacksburg, VA 24061, USA}

\author{Takeshi Yanai}
\affiliation{Institute of Transformative Bio-Molecules (WPI-ITbM), Nagoya University, Furocho, Chikusa Ward, Nagoya, Aichi 464-8601, Japan}

\author{Edward~F.~Valeev}
\email{efv@vt.edu}
\affiliation{Department of Chemistry, Virginia Tech, Blacksburg, VA 24061, USA}

\date{\today}

\maketitle %

In Ref. \onlinecite{VRG:yanai:2012:JCP} Yanai and Shiozaki presented a formalism for regularizing the Coulomb Hamiltonian by approximate similarity transformation (transcorrelation) with explicitly correlated geminals. The {\em a priori} inclusion of the explicitly correlated terms into the Hamiltonian, rather than into the wave function/operator, is formally appealing; combined with robust reduction of the basis set error and the fact that the transformed Hamiltonian  only contains 2-particle interactions (albeit, unlike the regular Coulomb interactions, they are nonlocal in real space) attracted several research groups\cite{ ets:sharma:2014, ets:yanai:2014,VRG:kersten:2016:JCP,VRG:motta:2020:PCCP,VRG:kumar:2022:JCTC} to investigate the approach. The goal of this Comment is to identify and correct errors in the formalism/implementation reported in Ref. \onlinecite{VRG:yanai:2012:JCP} and discuss some aspects of that work that were not fully specified in the original publication. This Comment also provides reference numerical results for a simple system to ease future implementation of the approach by other researchers.

\begin{itemize}
\item Eq. (27) in Ref. \onlinecite{VRG:yanai:2012:JCP} contains spurious factor of 1/2; it should be omitted to obtain the correct expression. We discovered the error by comparing the manual implementation ({\bf I2}) of the formulas reported in \onlinecite{VRG:yanai:2012:JCP} (developed in the course of work reported in Ref. \onlinecite{VRG:motta:2020:PCCP}) against the automated implementation ({\bf I3}) of the operator algebra using version 2 of the \code{SeQuant} toolkit\cite{VRG:valeev:2022:sequant-2.0.0} that will be described elsewhere.
For the reference purposes the corrected expressions for the approximate transcorrelated (CT-F12) Hamiltonian  and its tensor elements are documented in the Supplemental Material (SM) for this article. 
\item The original computer implementation used to generate the numerical data reported in Ref. \onlinecite{VRG:yanai:2012:JCP}  ({\bf I1}) contained errors. Some (but not all) of these errors were discovered in the course of detailed analysis of the numerical differences between the two manual implementations, {\bf I1} and {\bf I2}. Note that these two implementations are completely independent (e.g., they even use different evaluation schemes for the AO integrals over correlation factor-containing kernels: Gauss-Rys quadrature used by {\bf I1}\cite{VRG:shiozaki:2009:CPL} and the Obara-Saika recurrence used by {\bf I2}\cite{VRG:valeev:2021:libint-2.7.0}). Numerical results produced by {\bf I2} (manual) and {\bf I3} (automated) agreed perfectly; these implementations are integrated into the \code{MPQC} framework\cite{VRG:peng:2020:JCP} (see git commit {\tt 4f19136fda66bd7cf06863629edbf4ce1508bf2d}) the first as a unit test and the latter as a end-user class, and thus both share its numerical technologies. Due to their deep integration in \code{MPQC} it is not possible at the moment to full document these implementations by standalone source code; they will become available as part of the next public release of \code{MPQC}. However, the non-factorized equations that are automatically generated as part of the {\bf I3} implementation can be produced and verified using a publicly-available version of \code{SeQuant}.\footnote{See \url{ https://github.com/ValeevGroup/SeQuant/blob/e8067c72c4d8e8d3b8ddbd9eea8775244aadd1fa/examples/uccf12/uccf12.cpp}.}
\item The application of the frozen core approximation in the CT-F12 framework was unfortunately not fully described in Ref. \onlinecite{VRG:yanai:2012:JCP}, hence we discuss this issue here. Since the Slater-type geminal parameters recommended for standard F12 calculations are appropriate only for valence-only computations, the frozen core approximation in the CT-F12 approach should be first introduced ({\bf a}) by excluding core orbitals from the geminal-generating orbitals (Eq. (10) of Ref. \onlinecite{VRG:yanai:2012:JCP}). It may be also reasonable to ({\bf b}) exclude the core orbitals from the  transcorrelated contributions to the Hamiltonian (Eqs. (17) and (20) of Ref. \onlinecite{VRG:yanai:2012:JCP}). Due to the use of the cumulant decomposition to approximate the 3-body operators in the CT-F12 method this leads to two further subvariants of this approach: ({\bf b1}) with core orbitals excluded from the 3-body terms in the non-approximated transformed Hamiltonian (see Eq. (S1) in SM) {\em before} invoking the cumulant decomposition, and ({\bf b2}) vice versa, with the cumulant decomposition preceding the core orbital exclusion.
This seemingly innocuous order reversal leads to the appearance of RDM elements involving core orbitals in the frozen-core CT-F12 formulation ({\bf a+b2}) but not in its ({\bf a+b1}) counterpart (Eqs. (S10) and (S8), respectively). In Refs. \onlinecite{VRG:motta:2020:PCCP,VRG:kumar:2022:JCTC} as well as in this work we used frozen core formulation ({\bf a}). In Ref. \onlinecite{VRG:yanai:2012:JCP} formulation ({\bf a+b1}) was used; in other words all occupied and OBS ranges in that work exclude the core orbitals. The differences between the three frozen-core formulations in practice may be small, but not negligible and the distinction is important for the purposes of reproducibility. Plausible arguments for all 3 frozen-core variants can be put forth, especially if one considers extensions of the CT-F12 approach including single-particle relaxations as explored recently in Ref. \onlinecite{VRG:kumar:2022:JCTC}. To keep our focus on the issues in Ref. \onlinecite{VRG:yanai:2012:JCP} we do not further investigate numerical differences between the three frozen-core approaches.
\end{itemize}

Table \ref{tab:ne-data} contains reference results for a neon atom obtained with the transformed Hamiltonian for a number of standard single-reference correlated methods. These results can be directly compared with the results from Table II of Yanai and Shiozaki. The same method nomenclature as in Ref. \onlinecite{VRG:yanai:2012:JCP} is used here, i.e., F12-$X$ refers to method $X$ using the CT-F12 Hamiltonian. All computations utilized aug-cc-pVXZ orbital basis sets\cite{VRG:dunning:1989:JCP,VRG:kendall:1992:JCP} and the matching aug-cc-pVXZ/OptRI bases\cite{VRG:yousaf:2008:JCP} for the  CABS\cite{VRG:valeev:2004:CPL} construction. The correlation factor, $1 - \exp(-\gamma r_{12})/\gamma$, with $\gamma = 1.5 a_0$, was not approximated by fitting to Gaussians (as is done traditionally\cite{VRG:may:2005:PCCPP,VRG:tew:2005:JCP}), i.e., integrals over the ``genuine'' factor were employed in all calculations as in Ref. \onlinecite{VRG:yanai:2012:JCP}. All calculations were performed with the developmental version of the \code{MPQC} software package\cite{VRG:peng:2020:JCP}. No density fitting approximation was used. Unlike the TCE-based coupled-cluster computations in Ref. \onlinecite{VRG:yanai:2012:JCP}, the 8-fold permutational symmetry was not enforced for the coupled-cluster computations.

\begin{table*}[tbp]
    \centering
    \begin{tabular}{ccdddl}
    \hline\hline
       Method & Source & \text{aug-cc-pVDZ} & \text{aug-cc-pVTZ} & \text{aug-cc-pVQZ} & \text{CBS} \\ \hline
       HF & this work & -128.496349731  & -128.533272825 & -128.543755937 & -128.5474\footnotemark[1] \\ \hline
       \multirow{2}{*}{F12-HF} & Ref. \citenum{VRG:yanai:2012:JCP} & -0.11148 & -0.04356 & -0.02025 & \multirow{3}{*}{$\Biggr\}$$0$\footnotemark[2]} \\
        & this work & -0.111555079 & -0.042845640 & -0.019939990 \\
        (MP2-)F12\footnotemark[3]
       & this work & -0.104682301 & -0.043083913 & -0.020967256   \\ \hline
       \multirow{2}{*}{F12-MP2} & Ref. \citenum{VRG:yanai:2012:JCP} & -0.31411 & -0.31171 & -0.31478 & \multirow{3}{*}{$\Biggr\}$$-0.3202$\footnotemark[4]}\\
       & this work & -0.301361902 & -0.308391143 & -0.313067546 \\
       MP2-F12 & this work & -0.311555810  & -0.315602818 & -0.318210062 \\ \hline
       \multirow{2}{*}{F12-CCSD} & Ref. \citenum{VRG:yanai:2012:JCP} & -0.31783 & -0.31356 & -0.31542 & \multirow{3}{*}{$\Biggr\}$$-0.3157$\footnotemark[5]} \\
        & this work & -0.30713026 & -0.310390088 & -0.313651909\\
        CCSD(2)$_{\overline{\mathrm{F12}}}$ & this work & -0.301489064 & -0.307734267 & -0.311958932 & \\ \hline
       \multirow{2}{*}{F12-CCSDT} & Ref. \citenum{VRG:yanai:2012:JCP} & -0.32055 & -0.31870 & -0.32126 & \multirow{3}{*}{$\Biggr\}$$-0.3220$\footnotemark[6]} \\
        & this work & -0.309736623 & -0.315500325 & -0.319487083 & \\
       CCSDT(2)$_{\overline{\mathrm{F12}}}$ & this work & -0.304635289 & -0.312921334 & -0.317828063 \\
       \hline\hline
    \end{tabular}
    \footnotetext[1]{Ref. \onlinecite{VRG:fischer:1963:CJP}}
    \footnotetext[2]{All F12 corrections become zero in the CBS limit.}
    \footnotetext[3]{$E$(MP2-F12) - $E$(MP2), i.e. the F12 correction to the MP2 energy.}
    \footnotetext[4]{Ref. \onlinecite{VRG:klopper:2002:JCP}}
    \footnotetext[5]{Ref. \onlinecite{VRG:valeev:2008:PCCPP}}
    \footnotetext[6]{The CBS CCSD energy from Ref. \onlinecite{VRG:valeev:2008:PCCPP} corrected for the difference between CCSDT-F12 and CCSD-F12 energies evaluated with the aug-cc-pV6Z basis from Ref. \cite{VRG:shiozaki:2009:JCP}.}
    \caption{Reference electronic energies ($E_\mathrm{h}$) of the neon atom obtained with conventional and transcorrelated Hamiltonians (to be compared to Table II in Ref. \onlinecite{VRG:yanai:2012:JCP}).}
    \label{tab:ne-data}
\end{table*}

The most significant conclusion from the data in \cref{tab:ne-data} is that the basis set convergence of the CT-F12 energies is {\em monotonic}, and similar to that of the traditional F12 counterparts. The origin of the troublesome {\em non-monotonic} basis set convergence of CT-F12 energies reported in Ref. \onlinecite{VRG:yanai:2012:JCP} should be largely attributed to the implementation errors. The reported F12-CC energies for Ne also have smaller errors than their conventional explicitly correlated coupled-cluster (CC-F12) counterparts. These findings suggest that the CT-F12 approach might be a good candidate for reducing the basis set error of the high-order CC methods, perhaps better than the traditional high-order CC-F12 approaches.\cite{VRG:shiozaki:2009:JCP} Further investigation along these lines will be reported shortly elsewhere.

This work underscores the importance of automation of all steps involved in the development of many-body electronic structure methods, no matter how simple. While automation does not solve all problems, it helps to reduce the vast room for formal and technical mistakes in developing such methods.

\section*{Supplemental Material}

See supplemental material for the reference formulas of the CT-F12 Hamiltonian and detailed discussion of its frozen core variants.

\begin{acknowledgements}
This work was supported by the US Department of Energy, Office of Science, via Award DE-SC0019374.
\end{acknowledgements}

\bibliography{vrgrefs, extrasource}

\end{document}


\newtagform{suppeqtag}{(S}{)}
\usetagform{suppeqtag}

\title{Supplemental Material for 'Comment on ``Canonical transcorrelated theory with projected Slater-type geminals''  [J. Chem. Phys. 136, 084107 (2012)]'} %

\author{Conner Masteran}
\affiliation{Department of Chemistry, Virginia Tech, Blacksburg, Virginia 24061, USA}

\author{Ashutosh Kumar}
\affiliation{Theoretical Division, Los Alamos National Laboratory, Los Alamos, New Mexico 87545, United States}

\author{Nakul Teke}
\affiliation{Department of Chemistry, Virginia Tech, Blacksburg, VA 24061, USA}

\author{Bimal Gaudel}
\affiliation{Department of Chemistry, Virginia Tech, Blacksburg, VA 24061, USA}

\author{Takeshi Yanai}
\affiliation{Institute of Transformative Bio-Molecules (WPI-ITbM), Nagoya University, Furocho, Chikusa Ward, Nagoya, Aichi 464-8601, Japan}

\author{Edward~G.~Valeev}
\email{efv@vt.edu}
\affiliation{Department of Chemistry, Virginia Tech, Blacksburg, Virginia 24061, USA}

\date{\today}

\maketitle %

Here we document the reference expressions for the transcorrelated Hamiltonian described in \onlinecite{VRG:yanai:2012:JCP}.
All notation and definitions used here exactly match Ref. \onlinecite{VRG:yanai:2012:JCP}, with one extension: primed indices $p'_1, p'_2, \dots$ denote orbital basis set (OBS) that includes core orbitals even if the frozen-core (valence-only) approximation is invoked. As discussed in the main manuscript, the OBS range in Ref. \onlinecite{VRG:yanai:2012:JCP} excludes core orbitals in the valence-only (frozen-core) computations. Both here and in Ref. \onlinecite{VRG:yanai:2012:JCP} the definition of geminal operators involves valence occupied orbitals only (i.e., the geminal excitations do not occur from the core orbitals).

A complete expression for the transcorrelated Hamiltonian
obtained {\em without the use of the cumulant decomposition} is:
\begin{align}
\label{eq:H}
\hat{H} + [\hat{H}, \hat{A}^\mathrm{F12}] + \frac{1}{2} [[\hat{F}, \hat{A}^\mathrm{F12}], \hat{A}^\mathrm{F12}] & = {{{\frac{1}{2}}}{g^{{{p^\prime}_1}{{p^\prime}_2}}_{{\alpha_1}{\alpha_2}}}{G^{{\alpha_1}{\alpha_2}}_{{i_1}{i_2}}}{\hat{E}^{{i_1}{i_2}}_{{{p^\prime}_1}{{p^\prime}_2}}}} + {{h^{{{p^\prime}_1}}_{{x_1}}}{G^{{a_1}{x_1}}_{{i_1}{i_2}}}{\hat{E}^{{i_1}{i_2}}_{{a_1}{{p^\prime}_1}}}} \nonumber \\
  & + {{{\frac{1}{2}}}{g^{{\alpha_1}{\alpha_2}}_{{{p^\prime}_1}{{p^\prime}_2}}}{G^{{i_1}{i_2}}_{{\alpha_1}{\alpha_2}}}{\hat{E}^{{{p^\prime}_1}{{p^\prime}_2}}_{{i_1}{i_2}}}} + {{h^{{x_1}}_{{{p^\prime}_1}}}{G^{{i_1}{i_2}}_{{a_1}{x_1}}}{\hat{E}^{{a_1}{{p^\prime}_1}}_{{i_1}{i_2}}}} \nonumber \\
  & + {{g^{{{p^\prime}_1}{x_1}}_{{{p^\prime}_2}{{p^\prime}_3}}}{G^{{i_2}{i_1}}_{{a_1}{x_1}}}{\hat{E}^{{{p^\prime}_3}{a_1}{{p^\prime}_2}}_{{i_1}{i_2}{{p^\prime}_1}}}} + {{g^{{{p^\prime}_1}{{p^\prime}_3}}_{{{p^\prime}_2}{x_1}}}{G^{{x_1}{a_1}}_{{i_1}{i_2}}}{\hat{E}^{{i_2}{{p^\prime}_2}{i_1}}_{{a_1}{{p^\prime}_1}{{p^\prime}_3}}}}  \nonumber \\
  & + {{f^{{x_2}}_{{x_1}}}{G^{{a_1}{x_1}}_{{i_1}{i_2}}}{G^{{i_4}{i_3}}_{{a_2}{x_2}}}{\hat{E}^{{i_2}{a_2}{i_1}}_{{i_3}{i_4}{a_1}}}} - {{{\frac{1}{2}}}{f^{{i_1}}_{{{p^\prime}_1}}}{G^{{\alpha_1}{\alpha_2}}_{{i_1}{i_2}}}{G^{{i_3}{i_4}}_{{\alpha_1}{\alpha_2}}}{\hat{E}^{{{p^\prime}_1}{i_2}}_{{i_3}{i_4}}}} \nonumber \\
  & - {{{\frac{1}{2}}}{f^{{{p^\prime}_1}}_{{i_1}}}{G^{{\alpha_1}{\alpha_2}}_{{i_2}{i_3}}}{G^{{i_4}{i_1}}_{{\alpha_1}{\alpha_2}}}{\hat{E}^{{i_2}{i_3}}_{{i_4}{{p^\prime}_1}}}} + {{{\frac{1}{2}}}{f^{{\alpha_1}}_{{{p^\prime}_1}}}{G^{{a_1}{x_1}}_{{i_1}{i_2}}}{G^{{i_3}{i_4}}_{{x_1}{\alpha_1}}}{\hat{E}^{{i_2}{{p^\prime}_1}{i_1}}_{{i_3}{i_4}{a_1}}}} \nonumber \\
  & + {{{\frac{1}{2}}}{f^{{{p^\prime}_1}}_{{\alpha_1}}}{G^{{i_1}{i_2}}_{{a_1}{x_1}}}{G^{{x_1}{\alpha_1}}_{{i_3}{i_4}}}{\hat{E}^{{a_1}{i_3}{i_4}}_{{i_1}{i_2}{{p^\prime}_1}}}} - {{{\frac{1}{2}}}{f^{{{p^\prime}_1}}_{{i_2}}}{G^{{i_1}{i_2}}_{{a_1}{x_1}}}{G^{{a_2}{x_1}}_{{i_3}{i_4}}}{\hat{E}^{{a_1}{i_3}{i_4}}_{{i_1}{a_2}{{p^\prime}_1}}}} \nonumber \\
  & - {{{\frac{1}{2}}}{f^{{{p^\prime}_1}}_{{i_2}}}{G^{{a_1}{x_1}}_{{i_1}{i_3}}}{G^{{i_2}{i_4}}_{{a_2}{x_1}}}{\hat{E}^{{i_3}{i_1}{a_2}}_{{i_4}{a_1}{{p^\prime}_1}}}} - {{{\frac{1}{2}}}{f^{{i_2}}_{{{p^\prime}_1}}}{G^{{a_1}{x_1}}_{{i_1}{i_2}}}{G^{{i_3}{i_4}}_{{a_2}{x_1}}}{\hat{E}^{{a_2}{{p^\prime}_1}{i_1}}_{{i_3}{i_4}{a_1}}}} \nonumber \\
  & - {{{\frac{1}{2}}}{f^{{i_2}}_{{{p^\prime}_1}}}{G^{{i_1}{i_3}}_{{a_1}{x_1}}}{G^{{a_2}{x_1}}_{{i_2}{i_4}}}{\hat{E}^{{a_1}{i_4}{{p^\prime}_1}}_{{i_1}{i_3}{a_2}}}} + {{f^{{\alpha_2}}_{{\alpha_1}}}{G^{{\alpha_1}{\alpha_3}}_{{i_1}{i_2}}}{G^{{i_4}{i_3}}_{{\alpha_2}{\alpha_3}}}{\hat{E}^{{i_2}{i_1}}_{{i_3}{i_4}}}}.
\end{align}
Note the appearance of the core-including OBS indices ($p'$) in this expression.
These expressions are identical to the spin-orbital expressions modulo the replacement of the spin-free normal-ordered (with respect to genuine vacuum) replacement operators $\hat{E}$ with the spin-orbital normal-ordered operators ($a$ in the notation of Ref. \onlinecite{VRG:kutzelnigg:1982:JCP}), and can be used as to obtain cumulant-approximated transformed Hamiltonian expressions applicable to relativistic Hamiltonians and/or odd numbers of electrons.

Use of the spin-free cumulant-based approximation to the 3-body operator components produces the approximate transcorrelated Hamiltonian (Eq. (9) in Ref. \onlinecite{VRG:yanai:2012:JCP}):
\begin{align}
\label{eq:H-approx}
\hat{\bar{H}}^\mathrm{F12} \equiv \hat{H} + [\hat{H}, \hat{A}^\mathrm{F12}]_{1,2} + \frac{1}{2} [[\hat{F}, \hat{A}^\mathrm{F12}]_{1,2}, \hat{A}^\mathrm{F12}]_{1,2} = \bar{h}^{p^\prime}_{q^\prime} \hat{E}^{q^\prime}_{p^\prime} + \frac{1}{2}\bar{g}^{p^\prime r^\prime}_{q^\prime s^\prime } \hat{E}^{q^\prime s^\prime}_{p^\prime r^\prime} 
\end{align}
Expressions for the 1- and 2-body matrix elements of the approximate transcorrelated Hamiltonian (Eqs. (14-15) in Ref. \onlinecite{VRG:yanai:2012:JCP}, respectively) are:
\begin{align}
\label{eq:h}
\bar{h}_{q^\prime}^{p^\prime} = 
  &  {{h_{q^\prime}^{p^\prime}}} - {{{\frac{1}{2}}}{D^{{{p^\prime}_3}}_{{{p^\prime}_2}}}{D^{{{p^\prime}_1}}_{{i_1}}}{G^{{i_1}{i_2}}_{{a_1}{x_1}}}{g^{{{p^\prime}_2}{x_1}}_{{{p^\prime}_1}{{p^\prime}_3}}}{\delta^{{a_1}}_{q^\prime}}{\delta^{p^\prime}_{{i_2}}}} \nonumber \\
  & - {{{2}}{D^{{{p^\prime}_2}}_{{{p^\prime}_3}}}{D^{{{p^\prime}_1}}_{{i_1}}}{G^{{i_2}{i_1}}_{{a_1}{x_1}}}{g^{{x_1}{{p^\prime}_3}}_{{{p^\prime}_1}{{p^\prime}_2}}}{\delta^{{a_1}}_{q^\prime}}{\delta^{p^\prime}_{{i_2}}}} + {{D^{{{p^\prime}_2}}_{{{p^\prime}_3}}}{D^{{i_1}}_{{{p^\prime}_1}}}{G^{{x_1}{a_1}}_{{i_1}{i_2}}}{g^{{{p^\prime}_1}{{p^\prime}_3}}_{{{p^\prime}_2}{x_1}}}{\delta^{{i_2}}_{q^\prime}}{\delta^{p^\prime}_{{a_1}}}} \nonumber \\
  & - {{{\frac{1}{2}}}{D^{{{p^\prime}_2}}_{{i_2}}}{D^{{{p^\prime}_1}}_{{i_1}}}{G^{{i_1}{i_2}}_{{a_1}{x_1}}}{g^{{x_1}{{p^\prime}_3}}_{{{p^\prime}_1}{{p^\prime}_2}}}{\delta^{{a_1}}_{q^\prime}}{\delta^{p^\prime}_{{{p^\prime}_3}}}} - {{{\frac{1}{2}}}{D^{{i_2}}_{{{p^\prime}_2}}}{D^{{i_1}}_{{{p^\prime}_1}}}{G^{{a_1}{x_1}}_{{i_1}{i_2}}}{g^{{{p^\prime}_2}{{p^\prime}_1}}_{{{p^\prime}_3}{x_1}}}{\delta^{{{p^\prime}_3}}_{q^\prime}}{\delta^{p^\prime}_{{a_1}}}} \nonumber \\
  & + {{D^{{i_2}}_{{{p^\prime}_2}}}{D^{{i_1}}_{{{p^\prime}_1}}}{G^{{x_1}{a_1}}_{{i_1}{i_2}}}{g^{{{p^\prime}_2}{{p^\prime}_1}}_{{{p^\prime}_3}{x_1}}}{\delta^{{{p^\prime}_3}}_{q^\prime}}{\delta^{p^\prime}_{{a_1}}}} - {{{\frac{1}{2}}}{D^{{{p^\prime}_2}}_{{{p^\prime}_3}}}{D^{{i_1}}_{{{p^\prime}_1}}}{G^{{a_1}{x_1}}_{{i_1}{i_2}}}{g^{{{p^\prime}_1}{{p^\prime}_3}}_{{{p^\prime}_2}{x_1}}}{\delta^{{i_2}}_{q^\prime}}{\delta^{p^\prime}_{{a_1}}}} \nonumber \\
  & + {{D^{{i_1}}_{{{p^\prime}_3}}}{D^{{{p^\prime}_1}}_{{{p^\prime}_2}}}{G^{{a_1}{x_1}}_{{i_1}{i_2}}}{g^{{{p^\prime}_2}{{p^\prime}_3}}_{{{p^\prime}_1}{x_1}}}{\delta^{{i_2}}_{q^\prime}}{\delta^{p^\prime}_{{a_1}}}} + {{D^{{{p^\prime}_2}}_{{i_2}}}{D^{{{p^\prime}_1}}_{{i_1}}}{G^{{i_2}{i_1}}_{{a_1}{x_1}}}{g^{{x_1}{{p^\prime}_3}}_{{{p^\prime}_1}{{p^\prime}_2}}}{\delta^{{a_1}}_{q^\prime}}{\delta^{p^\prime}_{{{p^\prime}_3}}}} \nonumber \\
  & + {{D^{{{p^\prime}_2}}_{{{p^\prime}_3}}}{D^{{{p^\prime}_1}}_{{i_1}}}{G^{{i_1}{i_2}}_{{a_1}{x_1}}}{g^{{x_1}{{p^\prime}_3}}_{{{p^\prime}_1}{{p^\prime}_2}}}{\delta^{{a_1}}_{q^\prime}}{\delta^{p^\prime}_{{i_2}}}} - {{{2}}{D^{{i_1}}_{{{p^\prime}_3}}}{D^{{{p^\prime}_1}}_{{{p^\prime}_2}}}{G^{{x_1}{a_1}}_{{i_1}{i_2}}}{g^{{{p^\prime}_2}{{p^\prime}_3}}_{{{p^\prime}_1}{x_1}}}{\delta^{{i_2}}_{q^\prime}}{\delta^{p^\prime}_{{a_1}}}} \nonumber \\
  & + {{D^{{{p^\prime}_3}}_{{{p^\prime}_2}}}{D^{{{p^\prime}_1}}_{{i_1}}}{G^{{i_2}{i_1}}_{{a_1}{x_1}}}{g^{{{p^\prime}_2}{x_1}}_{{{p^\prime}_1}{{p^\prime}_3}}}{\delta^{{a_1}}_{q^\prime}}{\delta^{p^\prime}_{{i_2}}}} - {{{\frac{1}{2}}}{D^{{{p^\prime}_1}{i_1}}_{{{p^\prime}_2}{{p^\prime}_3}}}{G^{{a_1}{x_1}}_{{i_1}{i_2}}}{g^{{{p^\prime}_2}{{p^\prime}_3}}_{{{p^\prime}_1}{x_1}}}{\delta^{{i_2}}_{q^\prime}}{\delta^{p^\prime}_{{a_1}}}} \nonumber \\
  & + {{D^{{{p^\prime}_1}{{p^\prime}_2}}_{{i_1}{{p^\prime}_3}}}{G^{{i_2}{i_1}}_{{a_1}{x_1}}}{g^{{x_1}{{p^\prime}_3}}_{{{p^\prime}_1}{{p^\prime}_2}}}{\delta^{{a_1}}_{q^\prime}}{\delta^{p^\prime}_{{i_2}}}} - {{{\frac{1}{2}}}{D^{{{p^\prime}_1}{{p^\prime}_2}}_{{i_1}{i_2}}}{G^{{i_2}{i_1}}_{{a_1}{x_1}}}{g^{{x_1}{{p^\prime}_3}}_{{{p^\prime}_1}{{p^\prime}_2}}}{\delta^{{a_1}}_{q^\prime}}{\delta^{p^\prime}_{{{p^\prime}_3}}}} \nonumber \\
  & - {{{\frac{1}{2}}}{D^{{{p^\prime}_1}{{p^\prime}_2}}_{{i_1}{{p^\prime}_3}}}{G^{{i_1}{i_2}}_{{a_1}{x_1}}}{g^{{x_1}{{p^\prime}_3}}_{{{p^\prime}_1}{{p^\prime}_2}}}{\delta^{{a_1}}_{q^\prime}}{\delta^{p^\prime}_{{i_2}}}} - {{{\frac{1}{2}}}{D^{{i_1}{i_2}}_{{{p^\prime}_1}{{p^\prime}_2}}}{G^{{x_1}{a_1}}_{{i_1}{i_2}}}{g^{{{p^\prime}_2}{{p^\prime}_1}}_{{{p^\prime}_3}{x_1}}}{\delta^{{{p^\prime}_3}}_{q^\prime}}{\delta^{p^\prime}_{{a_1}}}} \nonumber \\
  & + {{D^{{{p^\prime}_1}{i_1}}_{{{p^\prime}_2}{{p^\prime}_3}}}{G^{{x_1}{a_1}}_{{i_1}{i_2}}}{g^{{{p^\prime}_2}{{p^\prime}_3}}_{{{p^\prime}_1}{x_1}}}{\delta^{{i_2}}_{q^\prime}}{\delta^{p^\prime}_{{a_1}}}} \nonumber \\
  & - {{{\frac{1}{4}}}{G^{{i_4}{i_2}}_{{a_3}{x_1}}}{G^{{a_1}{x_1}}_{{i_1}{i_3}}}{f^{{a_2}}_{{a_1}}}{D^{{i_3}}_{{i_4}}}{D^{{i_1}}_{{i_2}}}{\delta^{{a_3}}_{q^\prime}}{\delta^{p^\prime}_{{a_2}}}} - {{{\frac{1}{2}}}{G^{{i_2}{i_4}}_{{a_2}{x_1}}}{G^{{x_1}{a_1}}_{{i_3}{i_5}}}{f^{{i_5}}_{{i_1}}}{D^{{i_3}}_{{i_4}}}{D^{{i_1}}_{{i_2}}}{\delta^{{a_2}}_{q^\prime}}{\delta^{p^\prime}_{{a_1}}}} \nonumber \\
  & - {{{\frac{1}{2}}}{G^{{i_4}{i_2}}_{{a_2}{x_1}}}{G^{{a_1}{x_1}}_{{i_3}{i_5}}}{f^{{i_5}}_{{i_1}}}{D^{{i_3}}_{{i_4}}}{D^{{i_1}}_{{i_2}}}{\delta^{{a_2}}_{q^\prime}}{\delta^{p^\prime}_{{a_1}}}} - {{{\frac{1}{4}}}{G^{{i_2}{i_4}}_{{a_1}{x_1}}}{G^{{x_1}{a_3}}_{{i_1}{i_3}}}{f^{{a_1}}_{{a_2}}}{D^{{i_3}}_{{i_4}}}{D^{{i_1}}_{{i_2}}}{\delta^{{a_2}}_{q^\prime}}{\delta^{p^\prime}_{{a_3}}}} \nonumber \\
  & + {{{\frac{1}{2}}}{G^{{i_2}{i_4}}_{{a_3}{x_1}}}{G^{{a_1}{x_1}}_{{i_1}{i_3}}}{f^{{a_2}}_{{a_1}}}{D^{{i_3}}_{{i_4}}}{D^{{i_1}}_{{i_2}}}{\delta^{{a_3}}_{q^\prime}}{\delta^{p^\prime}_{{a_2}}}} + {{{\frac{1}{4}}}{G^{{i_2}{i_5}}_{{a_2}{x_1}}}{G^{{x_1}{a_1}}_{{i_1}{i_4}}}{f^{{i_3}}_{{i_5}}}{D^{{i_4}}_{{i_3}}}{D^{{i_1}}_{{i_2}}}{\delta^{{a_2}}_{q^\prime}}{\delta^{p^\prime}_{{a_1}}}} \nonumber \\
  & - {{{\frac{1}{2}}}{G^{{i_2}{i_5}}_{{a_2}{x_1}}}{G^{{a_1}{x_1}}_{{i_1}{i_4}}}{f^{{i_3}}_{{i_5}}}{D^{{i_4}}_{{i_3}}}{D^{{i_1}}_{{i_2}}}{\delta^{{a_2}}_{q^\prime}}{\delta^{p^\prime}_{{a_1}}}} + {{{\frac{1}{2}}}{G^{{i_2}{i_4}}_{{a_1}{x_1}}}{G^{{a_3}{x_1}}_{{i_1}{i_3}}}{f^{{a_1}}_{{a_2}}}{D^{{i_3}}_{{i_4}}}{D^{{i_1}}_{{i_2}}}{\delta^{{a_2}}_{q^\prime}}{\delta^{p^\prime}_{{a_3}}}} \nonumber \\
  & + {{{\frac{1}{4}}}{G^{{i_4}{i_2}}_{{a_2}{x_1}}}{G^{{x_1}{a_1}}_{{i_3}{i_5}}}{f^{{i_5}}_{{i_1}}}{D^{{i_3}}_{{i_4}}}{D^{{i_1}}_{{i_2}}}{\delta^{{a_2}}_{q^\prime}}{\delta^{p^\prime}_{{a_1}}}} - {{{\frac{1}{2}}}{G^{{i_5}{i_2}}_{{a_2}{x_1}}}{G^{{x_1}{a_1}}_{{i_1}{i_4}}}{f^{{i_3}}_{{i_5}}}{D^{{i_4}}_{{i_3}}}{D^{{i_1}}_{{i_2}}}{\delta^{{a_2}}_{q^\prime}}{\delta^{p^\prime}_{{a_1}}}} \nonumber \\
  & + {{{\frac{1}{4}}}{G^{{i_2}{i_4}}_{{a_2}{x_1}}}{G^{{a_1}{x_1}}_{{i_3}{i_5}}}{f^{{i_5}}_{{i_1}}}{D^{{i_3}}_{{i_4}}}{D^{{i_1}}_{{i_2}}}{\delta^{{a_2}}_{q^\prime}}{\delta^{p^\prime}_{{a_1}}}} + {{G^{{i_4}{i_2}}_{{a_2}{x_2}}}{G^{{x_1}{a_1}}_{{i_1}{i_3}}}{f^{{x_2}}_{{x_1}}}{D^{{i_3}}_{{i_4}}}{D^{{i_1}}_{{i_2}}}{\delta^{{a_2}}_{q^\prime}}{\delta^{p^\prime}_{{a_1}}}} \nonumber \\
  & - {{{\frac{1}{2}}}{G^{{i_4}{i_2}}_{{a_2}{x_2}}}{G^{{a_1}{x_1}}_{{i_1}{i_3}}}{f^{{x_2}}_{{x_1}}}{D^{{i_3}}_{{i_4}}}{D^{{i_1}}_{{i_2}}}{\delta^{{a_2}}_{q^\prime}}{\delta^{p^\prime}_{{a_1}}}} + {{{\frac{1}{4}}}{G^{{i_5}{i_2}}_{{a_2}{x_1}}}{G^{{a_1}{x_1}}_{{i_1}{i_4}}}{f^{{i_3}}_{{i_5}}}{D^{{i_4}}_{{i_3}}}{D^{{i_1}}_{{i_2}}}{\delta^{{a_2}}_{q^\prime}}{\delta^{p^\prime}_{{a_1}}}} \nonumber \\
  & - {{{\frac{1}{2}}}{G^{{i_4}{i_2}}_{{a_2}{x_2}}}{G^{{x_1}{a_1}}_{{i_1}{i_3}}}{f^{{x_2}}_{{x_1}}}{D^{{i_1}{i_3}}_{{i_2}{i_4}}}{\delta^{{a_2}}_{q^\prime}}{\delta^{p^\prime}_{{a_1}}}} - {{{\frac{1}{4}}}{G^{{i_2}{i_4}}_{{a_3}{x_1}}}{G^{{a_1}{x_1}}_{{i_1}{i_3}}}{f^{{a_2}}_{{a_1}}}{D^{{i_1}{i_3}}_{{i_2}{i_4}}}{\delta^{{a_3}}_{q^\prime}}{\delta^{p^\prime}_{{a_2}}}} \nonumber \\
  & - {{{\frac{1}{4}}}{G^{{i_2}{i_4}}_{{a_1}{x_1}}}{G^{{a_3}{x_1}}_{{i_1}{i_3}}}{f^{{a_1}}_{{a_2}}}{D^{{i_1}{i_3}}_{{i_2}{i_4}}}{\delta^{{a_2}}_{q^\prime}}{\delta^{p^\prime}_{{a_3}}}} + {{{\frac{1}{4}}}{G^{{i_5}{i_2}}_{{a_2}{x_1}}}{G^{{x_1}{a_1}}_{{i_1}{i_4}}}{f^{{i_3}}_{{i_5}}}{D^{{i_1}{i_4}}_{{i_2}{i_3}}}{\delta^{{a_2}}_{q^\prime}}{\delta^{p^\prime}_{{a_1}}}} \nonumber \\
  & + {{{\frac{1}{4}}}{G^{{i_2}{i_4}}_{{a_2}{x_1}}}{G^{{x_1}{a_1}}_{{i_3}{i_5}}}{f^{{i_5}}_{{i_1}}}{D^{{i_1}{i_3}}_{{i_2}{i_4}}}{\delta^{{a_2}}_{q^\prime}}{\delta^{p^\prime}_{{a_1}}}} + {{{\frac{1}{4}}}{G^{{i_4}{i_2}}_{{a_2}{x_1}}}{G^{{a_1}{x_1}}_{{i_3}{i_5}}}{f^{{i_5}}_{{i_1}}}{D^{{i_1}{i_3}}_{{i_2}{i_4}}}{\delta^{{a_2}}_{q^\prime}}{\delta^{p^\prime}_{{a_1}}}} \nonumber \\
  & + {{{\frac{1}{4}}}{G^{{i_2}{i_5}}_{{a_2}{x_1}}}{G^{{a_1}{x_1}}_{{i_1}{i_4}}}{f^{{i_3}}_{{i_5}}}{D^{{i_1}{i_4}}_{{i_2}{i_3}}}{\delta^{{a_2}}_{q^\prime}}{\delta^{p^\prime}_{{a_1}}}}
  \end{align}
  \begin{align}
  \label{eq:g}
{{\bar{g}^{p^\prime q^\prime}_{r^\prime s^\prime}}} = 
  & \hat{S} (  {{g^{p^\prime q^\prime }_{r^\prime s^\prime}}} - {{D^{{i_1}}_{{{p^\prime}_1}}}{G^{{x_1}{a_1}}_{{i_1}{i_2}}}{g^{{{p^\prime}_1}{{p^\prime}_3}}_{{{p^\prime}_2}{x_1}}}{\delta^{{i_2}}_{r^\prime}}{\delta^{{{p^\prime}_2}}_{s^\prime}}{\delta^{p^\prime}_{{a_1}}}{\delta^{q^\prime}_{{{p^\prime}_3}}}} \nonumber \\
  & - {{D^{{{p^\prime}_1}}_{{i_1}}}{G^{{i_1}{i_2}}_{{a_1}{x_1}}}{g^{{{p^\prime}_3}{x_1}}_{{{p^\prime}_1}{{p^\prime}_2}}}{\delta^{{{p^\prime}_2}}_{r^\prime}}{\delta^{{a_1}}_{s^\prime}}{\delta^{p^\prime}_{{i_2}}}{\delta^{q^\prime}_{{{p^\prime}_3}}}} - {{D^{{{p^\prime}_1}}_{{i_1}}}{G^{{i_2}{i_1}}_{{a_1}{x_1}}}{g^{{{p^\prime}_3}{x_1}}_{{{p^\prime}_1}{{p^\prime}_2}}}{\delta^{{{p^\prime}_2}}_{r^\prime}}{\delta^{{a_1}}_{s^\prime}}{\delta^{p^\prime}_{{{p^\prime}_3}}}{\delta^{q^\prime}_{{i_2}}}} \nonumber \\
  & + {{{2}}{D^{{{p^\prime}_2}}_{{{p^\prime}_1}}}{G^{{x_1}{a_1}}_{{i_1}{i_2}}}{g^{{{p^\prime}_1}{{p^\prime}_3}}_{{{p^\prime}_2}{x_1}}}{\delta^{{i_2}}_{r^\prime}}{\delta^{{i_1}}_{s^\prime}}{\delta^{p^\prime}_{{a_1}}}{\delta^{q^\prime}_{{{p^\prime}_3}}}} - {{D^{{{p^\prime}_1}}_{{{p^\prime}_2}}}{G^{{x_1}{a_1}}_{{i_1}{i_2}}}{g^{{{p^\prime}_3}{{p^\prime}_2}}_{{{p^\prime}_1}{x_1}}}{\delta^{{i_2}}_{r^\prime}}{\delta^{{i_1}}_{s^\prime}}{\delta^{p^\prime}_{{a_1}}}{\delta^{q^\prime}_{{{p^\prime}_3}}}} \nonumber \\
  & - {{D^{{i_1}}_{{{p^\prime}_1}}}{G^{{a_1}{x_1}}_{{i_1}{i_2}}}{g^{{{p^\prime}_2}{{p^\prime}_1}}_{{{p^\prime}_3}{x_1}}}{\delta^{{{p^\prime}_3}}_{r^\prime}}{\delta^{{i_2}}_{s^\prime}}{\delta^{p^\prime}_{{{p^\prime}_2}}}{\delta^{q^\prime}_{{a_1}}}} + {{{2}}{D^{{{p^\prime}_1}}_{{{p^\prime}_2}}}{G^{{i_1}{i_2}}_{{a_1}{x_1}}}{g^{{{p^\prime}_2}{x_1}}_{{{p^\prime}_1}{{p^\prime}_3}}}{\delta^{{{p^\prime}_3}}_{r^\prime}}{\delta^{{a_1}}_{s^\prime}}{\delta^{p^\prime}_{{i_2}}}{\delta^{q^\prime}_{{i_1}}}} \nonumber \\
  & + {{{2}}{D^{{i_1}}_{{{p^\prime}_1}}}{G^{{x_1}{a_1}}_{{i_1}{i_2}}}{g^{{{p^\prime}_2}{{p^\prime}_1}}_{{{p^\prime}_3}{x_1}}}{\delta^{{i_2}}_{r^\prime}}{\delta^{{{p^\prime}_3}}_{s^\prime}}{\delta^{p^\prime}_{{a_1}}}{\delta^{q^\prime}_{{{p^\prime}_2}}}} - {{D^{{{p^\prime}_1}}_{{i_1}}}{G^{{i_1}{i_2}}_{{a_1}{x_1}}}{g^{{x_1}{{p^\prime}_2}}_{{{p^\prime}_1}{{p^\prime}_3}}}{\delta^{{a_1}}_{r^\prime}}{\delta^{{{p^\prime}_3}}_{s^\prime}}{\delta^{p^\prime}_{{i_2}}}{\delta^{q^\prime}_{{{p^\prime}_2}}}} \nonumber \\
  & - {{D^{{{p^\prime}_2}}_{{{p^\prime}_1}}}{G^{{i_1}{i_2}}_{{a_1}{x_1}}}{g^{{x_1}{{p^\prime}_1}}_{{{p^\prime}_2}{{p^\prime}_3}}}{\delta^{{a_1}}_{r^\prime}}{\delta^{{{p^\prime}_3}}_{s^\prime}}{\delta^{p^\prime}_{{i_1}}}{\delta^{q^\prime}_{{i_2}}}} + {{{2}}{D^{{{p^\prime}_1}}_{{i_1}}}{G^{{i_2}{i_1}}_{{a_1}{x_1}}}{g^{{x_1}{{p^\prime}_2}}_{{{p^\prime}_1}{{p^\prime}_3}}}{\delta^{{a_1}}_{r^\prime}}{\delta^{{{p^\prime}_3}}_{s^\prime}}{\delta^{p^\prime}_{{i_2}}}{\delta^{q^\prime}_{{{p^\prime}_2}}}} \nonumber \\
  & - {{D^{{i_1}}_{{{p^\prime}_1}}}{G^{{a_1}{x_1}}_{{i_1}{i_2}}}{g^{{{p^\prime}_1}{{p^\prime}_3}}_{{{p^\prime}_2}{x_1}}}{\delta^{{{p^\prime}_2}}_{r^\prime}}{\delta^{{i_2}}_{s^\prime}}{\delta^{p^\prime}_{{a_1}}}{\delta^{q^\prime}_{{{p^\prime}_3}}}} + {{{2}}{G^{{a_1}{x_1}}_{{i_1}{i_2}}}{h^{{{p^\prime}_1}}_{{x_1}}}{\delta^{{i_1}}_{r^\prime}}{\delta^{{i_2}}_{s^\prime}}{\delta^{p^\prime}_{{a_1}}}{\delta^{q^\prime}_{{{p^\prime}_1}}}} \nonumber \\
  & + {{V^{{p^{\prime}_1}{p^{\prime}_2}}_{{i_1}{i_2}}}{\delta^{{i_1}}_{r^\prime}}{\delta^{{i_2}}_{s^\prime}}{\delta^{p^\prime}_{{p^{\prime}_1}}}{\delta^{q^\prime}_{{p^{\prime}_2}}}} + {{\tilde{V}^{{i_1}{i_2}}_{{p^{\prime}_1}{p^{\prime}_2}}}{\delta^{{p^{\prime}_1}}_{r^\prime}}{\delta^{{{p^\prime}_2}}_{s^\prime}}{\delta^{p^\prime}_{{i_1}}}{\delta^{q^\prime}_{{i_2}}}} \nonumber \\
  & + {{{2}}{G^{{i_1}{i_2}}_{{a_1}{x_1}}}{h^{{x_1}}_{{{p^\prime}_1}}}{\delta^{{a_1}}_{r^\prime}}{\delta^{{{p^\prime}_1}}_{s^\prime}}{\delta^{p^\prime}_{{i_1}}}{\delta^{q^\prime}_{{i_2}}}} \nonumber \\
  & + {{{\frac{1}{2}}}{G^{{i_5}{i_2}}_{{a_2}{x_1}}}{G^{{a_1}{x_1}}_{{i_3}{i_4}}}{f^{{i_3}}_{{i_1}}}{D^{{i_1}}_{{i_2}}}{\delta^{{a_2}}_{r^\prime}}{\delta^{{i_4}}_{s^\prime}}{\delta^{p^\prime}_{{i_5}}}{\delta^{q^\prime}_{{a_1}}}} + {{{\frac{1}{2}}}{G^{{i_5}{i_2}}_{{a_2}{x_1}}}{G^{{a_1}{x_1}}_{{i_1}{i_3}}}{f^{{i_3}}_{{i_4}}}{D^{{i_1}}_{{i_2}}}{\delta^{{a_2}}_{r^\prime}}{\delta^{{i_4}}_{s^\prime}}{\delta^{p^\prime}_{{i_5}}}{\delta^{q^\prime}_{{a_1}}}} \nonumber \\
  & - {{G^{{i_4}{i_2}}_{{a_2}{x_2}}}{G^{{a_1}{x_1}}_{{i_1}{i_3}}}{f^{{x_2}}_{{x_1}}}{D^{{i_1}}_{{i_2}}}{\delta^{{a_2}}_{r^\prime}}{\delta^{{i_3}}_{s^\prime}}{\delta^{p^\prime}_{{i_4}}}{\delta^{q^\prime}_{{a_1}}}} 
  - {{G^{{i_2}{i_4}}_{{a_2}{x_2}}}{G^{{x_1}{a_1}}_{{i_1}{i_3}}}{f^{{x_2}}_{{x_1}}}{D^{{i_1}}_{{i_2}}}{\delta^{{a_2}}_{r^\prime}}{\delta^{{i_3}}_{s^\prime}}{\delta^{p^\prime}_{{i_4}}}{\delta^{q^\prime}_{{a_1}}}} \nonumber \\
  & + {{{\frac{1}{2}}}{G^{{i_5}{i_3}}_{{a_2}{x_1}}}{G^{{a_1}{x_1}}_{{i_2}{i_4}}}{f^{{i_1}}_{{i_3}}}{D^{{i_2}}_{{i_1}}}{\delta^{{a_2}}_{r^\prime}}{\delta^{{i_4}}_{s^\prime}}{\delta^{p^\prime}_{{i_5}}}{\delta^{q^\prime}_{{a_1}}}} - {{{\frac{1}{2}}}{G^{{i_2}{i_4}}_{{a_3}{x_1}}}{G^{{a_1}{x_1}}_{{i_1}{i_3}}}{f^{{a_2}}_{{a_1}}}{D^{{i_1}}_{{i_2}}}{\delta^{{i_3}}_{r^\prime}}{\delta^{{a_3}}_{s^\prime}}{\delta^{p^\prime}_{{i_4}}}{\delta^{q^\prime}_{{a_2}}}} \nonumber \\
  & + {{{\frac{1}{2}}}{G^{{i_2}{i_3}}_{{a_2}{x_1}}}{G^{{a_1}{x_1}}_{{i_1}{i_5}}}{f^{{i_4}}_{{i_3}}}{D^{{i_1}}_{{i_2}}}{\delta^{{i_5}}_{r^\prime}}{\delta^{{a_2}}_{s^\prime}}{\delta^{p^\prime}_{{i_4}}}{\delta^{q^\prime}_{{a_1}}}} - {{{\frac{1}{2}}}{G^{{i_2}{i_4}}_{{a_1}{x_1}}}{G^{{a_3}{x_1}}_{{i_1}{i_3}}}{f^{{a_1}}_{{a_2}}}{D^{{i_1}}_{{i_2}}}{\delta^{{i_3}}_{r^\prime}}{\delta^{{a_2}}_{s^\prime}}{\delta^{p^\prime}_{{i_4}}}{\delta^{q^\prime}_{{a_3}}}} \nonumber \\
  & - {{{\frac{1}{2}}}{G^{{i_2}{i_4}}_{{a_1}{x_1}}}{G^{{x_1}{a_3}}_{{i_1}{i_3}}}{f^{{a_1}}_{{a_2}}}{D^{{i_1}}_{{i_2}}}{\delta^{{a_2}}_{r^\prime}}{\delta^{{i_3}}_{s^\prime}}{\delta^{p^\prime}_{{i_4}}}{\delta^{q^\prime}_{{a_3}}}} - {{G^{{i_5}{i_3}}_{{a_2}{x_1}}}{G^{{x_1}{a_1}}_{{i_2}{i_4}}}{f^{{i_1}}_{{i_3}}}{D^{{i_2}}_{{i_1}}}{\delta^{{a_2}}_{r^\prime}}{\delta^{{i_4}}_{s^\prime}}{\delta^{p^\prime}_{{i_5}}}{\delta^{q^\prime}_{{a_1}}}} \nonumber \\
  & + {{G^{{i_4}{i_2}}_{{a_1}{x_1}}}{G^{{x_1}{a_3}}_{{i_1}{i_3}}}{f^{{a_1}}_{{a_2}}}{D^{{i_1}}_{{i_2}}}{\delta^{{a_2}}_{r^\prime}}{\delta^{{i_3}}_{s^\prime}}{\delta^{p^\prime}_{{i_4}}}{\delta^{q^\prime}_{{a_3}}}} - {{{\frac{1}{2}}}{G^{{i_2}{i_4}}_{{a_3}{x_1}}}{G^{{x_1}{a_1}}_{{i_1}{i_3}}}{f^{{a_2}}_{{a_1}}}{D^{{i_1}}_{{i_2}}}{\delta^{{a_3}}_{r^\prime}}{\delta^{{i_3}}_{s^\prime}}{\delta^{p^\prime}_{{i_4}}}{\delta^{q^\prime}_{{a_2}}}} \nonumber \\
  & - {{{\frac{1}{2}}}{G^{{i_4}{i_2}}_{{a_1}{x_1}}}{G^{{a_3}{x_1}}_{{i_1}{i_3}}}{f^{{a_1}}_{{a_2}}}{D^{{i_1}}_{{i_2}}}{\delta^{{a_2}}_{r^\prime}}{\delta^{{i_3}}_{s^\prime}}{\delta^{p^\prime}_{{i_4}}}{\delta^{q^\prime}_{{a_3}}}} + {{{\frac{1}{2}}}{G^{{i_2}{i_5}}_{{a_2}{x_1}}}{G^{{a_1}{x_1}}_{{i_1}{i_3}}}{f^{{i_3}}_{{i_4}}}{D^{{i_1}}_{{i_2}}}{\delta^{{i_4}}_{r^\prime}}{\delta^{{a_2}}_{s^\prime}}{\delta^{p^\prime}_{{i_5}}}{\delta^{q^\prime}_{{a_1}}}} \nonumber \\
  & - {{G^{{i_5}{i_2}}_{{a_2}{x_1}}}{G^{{x_1}{a_1}}_{{i_3}{i_4}}}{f^{{i_3}}_{{i_1}}}{D^{{i_1}}_{{i_2}}}{\delta^{{a_2}}_{r^\prime}}{\delta^{{i_4}}_{s^\prime}}{\delta^{p^\prime}_{{i_5}}}{\delta^{q^\prime}_{{a_1}}}} - {{G^{{i_3}{i_2}}_{{a_2}{x_1}}}{G^{{x_1}{a_1}}_{{i_1}{i_5}}}{f^{{i_4}}_{{i_3}}}{D^{{i_1}}_{{i_2}}}{\delta^{{a_2}}_{r^\prime}}{\delta^{{i_5}}_{s^\prime}}{\delta^{p^\prime}_{{i_4}}}{\delta^{q^\prime}_{{a_1}}}} \nonumber \\
  & + {{{\frac{1}{2}}}{G^{{i_2}{i_5}}_{{a_2}{x_1}}}{G^{{x_1}{a_1}}_{{i_3}{i_4}}}{f^{{i_3}}_{{i_1}}}{D^{{i_1}}_{{i_2}}}{\delta^{{a_2}}_{r^\prime}}{\delta^{{i_4}}_{s^\prime}}{\delta^{p^\prime}_{{i_5}}}{\delta^{q^\prime}_{{a_1}}}} + {{{\frac{1}{2}}}{G^{{i_3}{i_5}}_{{a_2}{x_1}}}{G^{{a_1}{x_1}}_{{i_2}{i_4}}}{f^{{i_1}}_{{i_3}}}{D^{{i_2}}_{{i_1}}}{\delta^{{i_4}}_{r^\prime}}{\delta^{{a_2}}_{s^\prime}}{\delta^{p^\prime}_{{i_5}}}{\delta^{q^\prime}_{{a_1}}}} \nonumber \\
  & - {{{\frac{1}{2}}}{G^{{i_4}{i_2}}_{{a_3}{x_1}}}{G^{{a_1}{x_1}}_{{i_1}{i_3}}}{f^{{a_2}}_{{a_1}}}{D^{{i_1}}_{{i_2}}}{\delta^{{a_3}}_{r^\prime}}{\delta^{{i_3}}_{s^\prime}}{\delta^{p^\prime}_{{i_4}}}{\delta^{q^\prime}_{{a_2}}}} - {{G^{{i_5}{i_2}}_{{a_2}{x_1}}}{G^{{x_1}{a_1}}_{{i_1}{i_3}}}{f^{{i_3}}_{{i_4}}}{D^{{i_1}}_{{i_2}}}{\delta^{{a_2}}_{r^\prime}}{\delta^{{i_4}}_{s^\prime}}{\delta^{p^\prime}_{{i_5}}}{\delta^{q^\prime}_{{a_1}}}} \nonumber \\
  & + {{G^{{i_4}{i_2}}_{{a_3}{x_1}}}{G^{{x_1}{a_1}}_{{i_1}{i_3}}}{f^{{a_2}}_{{a_1}}}{D^{{i_1}}_{{i_2}}}{\delta^{{a_3}}_{r^\prime}}{\delta^{{i_3}}_{s^\prime}}{\delta^{p^\prime}_{{i_4}}}{\delta^{q^\prime}_{{a_2}}}} + {{{\frac{1}{2}}}{G^{{i_2}{i_5}}_{{a_2}{x_1}}}{G^{{a_1}{x_1}}_{{i_3}{i_4}}}{f^{{i_3}}_{{i_1}}}{D^{{i_1}}_{{i_2}}}{\delta^{{i_4}}_{r^\prime}}{\delta^{{a_2}}_{s^\prime}}{\delta^{p^\prime}_{{i_5}}}{\delta^{q^\prime}_{{a_1}}}} \nonumber \\
  & + {{{\frac{1}{2}}}{G^{{i_2}{i_3}}_{{a_2}{x_1}}}{G^{{x_1}{a_1}}_{{i_1}{i_5}}}{f^{{i_4}}_{{i_3}}}{D^{{i_1}}_{{i_2}}}{\delta^{{a_2}}_{r^\prime}}{\delta^{{i_5}}_{s^\prime}}{\delta^{p^\prime}_{{i_4}}}{\delta^{q^\prime}_{{a_1}}}} 
  + {{{\frac{1}{2}}}{G^{{i_3}{i_5}}_{{a_2}{x_1}}}{G^{{x_1}{a_1}}_{{i_2}{i_4}}}{f^{{i_1}}_{{i_3}}}{D^{{i_2}}_{{i_1}}}{\delta^{{a_2}}_{r^\prime}}{\delta^{{i_4}}_{s^\prime}}{\delta^{p^\prime}_{{i_5}}}{\delta^{q^\prime}_{{a_1}}}} \nonumber \\
  & + {{{\frac{1}{2}}}{G^{{i_3}{i_2}}_{{a_2}{x_1}}}{G^{{a_1}{x_1}}_{{i_1}{i_5}}}{f^{{i_4}}_{{i_3}}}{D^{{i_1}}_{{i_2}}}{\delta^{{a_2}}_{r^\prime}}{\delta^{{i_5}}_{s^\prime}}{\delta^{p^\prime}_{{i_4}}}{\delta^{q^\prime}_{{a_1}}}} + {{{2}}{G^{{i_4}{i_2}}_{{a_2}{x_2}}}{G^{{x_1}{a_1}}_{{i_1}{i_3}}}{f^{{x_2}}_{{x_1}}}{D^{{i_1}}_{{i_2}}}{\delta^{{a_2}}_{r^\prime}}{\delta^{{i_3}}_{s^\prime}}{\delta^{p^\prime}_{{i_4}}}{\delta^{q^\prime}_{{a_1}}}} \nonumber \\
  & - {{G^{{i_2}{i_4}}_{{a_2}{x_2}}}{G^{{a_1}{x_1}}_{{i_1}{i_3}}}{f^{{x_2}}_{{x_1}}}{D^{{i_1}}_{{i_2}}}{\delta^{{i_3}}_{r^\prime}}{\delta^{{a_2}}_{s^\prime}}{\delta^{p^\prime}_{{i_4}}}{\delta^{q^\prime}_{{a_1}}}} + {{{\frac{1}{2}}}{G^{{i_2}{i_5}}_{{a_2}{x_1}}}{G^{{x_1}{a_1}}_{{i_1}{i_3}}}{f^{{i_3}}_{{i_4}}}{D^{{i_1}}_{{i_2}}}{\delta^{{a_2}}_{r^\prime}}{\delta^{{i_4}}_{s^\prime}}{\delta^{p^\prime}_{{i_5}}}{\delta^{q^\prime}_{{a_1}}}} \nonumber \\
  & - {{X^{{i_3}{i_1}}_{{i_4}{i_5}}}{f^{{i_2}}_{{i_1}}}{\delta^{{i_5}}_{r^\prime}}{\delta^{{i_4}}_{s^\prime}}{\delta^{p^\prime}_{{i_2}}}{\delta^{q^\prime}_{{i_3}}}} - {{X^{{i_5}{i_3}}_{{i_1}{i_4}}}{f^{{i_1}}_{{i_2}}}{\delta^{{i_4}}_{r^\prime}}{\delta^{{i_2}}_{s^\prime}}{\delta^{p^\prime}_{{i_3}}}{\delta^{q^\prime}_{{i_5}}}} \nonumber \\
  & + {{B^{{i_2}{i_3}}_{{i_1}{i_4}}}{\delta^{{i_1}}_{r^\prime}}{\delta^{{i_4}}_{s^\prime}}{\delta^{p^\prime}_{{i_2}}}{\delta^{q^\prime}_{{i_3}}}}),
  \end{align}
where $\hat{S}$ is the particle symmetrization operator:
\begin{align}
\hat{S}o^{p^\prime q^\prime}_{r^\prime s^\prime} \equiv \frac{1}{2}(o^{p^\prime q^\prime}_{r^\prime s^\prime} + o^{q^\prime p^\prime}_{s^\prime r^\prime})
\end{align}
and where we distinguish the usual nonhermitian F12 intermediate $V$ (Eq. (23) in Ref. \onlinecite{VRG:yanai:2012:JCP}),
\begin{align}
{V_{{i_1}{i_2}}^{{p^{\prime}_1}{p^{\prime}_2}}} \equiv \, & g_{\alpha \beta}^{p^{\prime}_1 p^{\prime}_2} G_{i_1 i_2}^{\alpha \beta} ,
\end{align}
from its adjoint
\begin{align}
{\tilde{V}^{{i_1}{i_2}}_{{p^{\prime}_1}{p^{\prime}_2}}} \equiv \, & g^{\alpha \beta}_{p^{\prime}_1 p^{\prime}_2} G^{i_1 i_2}_{\alpha \beta} = \left( {V_{{i_1}{i_2}}^{{p^{\prime}_1}{p^{\prime}_2}}} \right)^* \neq {V^{{i_1}{i_2}}_{{p^{\prime}_1}{p^{\prime}_2}}}.
\end{align}
\Cref{eq:H-approx,eq:h,eq:g} can be verified using the open-source {\tt SeQuant} software (see \url{https://github.com/ValeevGroup/SeQuant/blob/e8067c72c4d8e8d3b8ddbd9eea8775244aadd1fa/examples/uccf12/uccf12.cpp}).

Note the appearance of the primed (core-including) OBS indices in \cref{eq:H-approx,eq:h,eq:g}. This is due to the exclusion of the core orbitals {\em only} in the definition of the geminal operator; this is referred to as frozen-core approach ({\bf a}) in the main text.
Ref. \onlinecite{VRG:yanai:2012:JCP} implemented approach ({\bf a+b1}) in which core orbitals are eliminated from the normal-ordered replacement operators in \cref{eq:H} {\em before} invoking the cumulant decomposition. As a concrete example consider $\hat{E}_{p^{\prime}_{3}a_{1}p^{\prime}_{2}}^{i_1 i_2 p^{\prime}_1}$ appearing in the 6th term on the right-hand side of \cref{eq:H}.
In approach ({\bf b1}) the initial elimination of the core orbitals (``fzc'') turns it into $\hat{E}_{p_{3}a_{1}p_{2}}^{i_1 i_2 p_1}$, whose subsequent cumulant decomposition (``cm'') yields
 \begin{align}
 \label{eq:E3-b1}
\hat{E}_{p^{\prime}_{3}a_{1}p^{\prime}_{2}}^{i_1 i_2 p^{\prime}_1} \overset{\rm {\bf b1}=fzc+cm}{\approx}& \bigg( - {{{\frac{1}{2}}}{D^{{i_2}}_{{p_3}}}{\hat{E}^{{p_1}{i_1}}_{{p_2}{a_1}}}} + {{{\frac{1}{2}}}{D^{{i_1}}_{{a_1}}}{D^{{p_1}{i_2}}_{{p_2}{p_3}}}} \nonumber \\
  & + {{D^{{p_1}{i_2}}_{{p_2}{a_1}}}{\hat{E}^{{i_1}}_{{p_3}}}} - {{{\frac{1}{2}}}{D^{{i_1}}_{{p_2}}}{\hat{E}^{{p_1}{i_2}}_{{p_3}{a_1}}}} + {{D^{{i_1}}_{{p_3}}}{\hat{E}^{{p_1}{i_2}}_{{p_2}{a_1}}}} \nonumber \\
  & - {{D^{{p_1}}_{{p_2}}}{D^{{i_1}{i_2}}_{{p_3}{a_1}}}} - {{{\frac{1}{2}}}{D^{{i_1}{p_1}}_{{p_3}{a_1}}}{\hat{E}^{{i_2}}_{{p_2}}}} - {{D^{{i_1}}_{{p_3}}}{D^{{p_1}{i_2}}_{{p_2}{a_1}}}} \nonumber \\
  & + {{{\frac{1}{2}}}{D^{{i_1}}_{{p_2}}}{D^{{p_1}{i_2}}_{{p_3}{a_1}}}} - {{{\frac{1}{2}}}{D^{{i_2}}_{{p_2}}}{\hat{E}^{{i_1}{p_1}}_{{p_3}{a_1}}}} - {{{\frac{1}{2}}}{D^{{p_1}{i_2}}_{{p_3}{a_1}}}{\hat{E}^{{i_1}}_{{p_2}}}} \nonumber \\
  & + {{{\frac{1}{2}}}{D^{{i_2}}_{{p_2}}}{D^{{i_1}{p_1}}_{{p_3}{a_1}}}} - {{{\frac{1}{2}}}{D^{{p_1}{i_1}}_{{p_2}{a_1}}}{\hat{E}^{{i_2}}_{{p_3}}}} - {{{\frac{1}{2}}}{D^{{i_2}{i_1}}_{{p_2}{p_3}}}{\hat{E}^{{p_1}}_{{a_1}}}} \nonumber \\
  & - {{{\frac{1}{2}}}{D^{{i_1}{i_2}}_{{p_2}{a_1}}}{\hat{E}^{{p_1}}_{{p_3}}}} - {{{\frac{1}{2}}}{D^{{p_1}}_{{a_1}}}{\hat{E}^{{i_2}{i_1}}_{{p_2}{p_3}}}} - {{{\frac{1}{2}}}{D^{{p_1}}_{{p_3}}}{\hat{E}^{{i_1}{i_2}}_{{p_2}{a_1}}}} \nonumber \\
  & + {{D^{{i_2}}_{{a_1}}}{\hat{E}^{{p_1}{i_1}}_{{p_2}{p_3}}}} + {{D^{{i_1}{i_2}}_{{p_3}{a_1}}}{\hat{E}^{{p_1}}_{{p_2}}}} - {{{\frac{1}{2}}}{D^{{i_1}}_{{a_1}}}{\hat{E}^{{p_1}{i_2}}_{{p_2}{p_3}}}} \nonumber \\
  & + {{{\frac{1}{2}}}{D^{{i_2}}_{{p_3}}}{D^{{p_1}{i_1}}_{{p_2}{a_1}}}} + {{D^{{p_1}}_{{p_2}}}{\hat{E}^{{i_1}{i_2}}_{{p_3}{a_1}}}} + {{{\frac{1}{2}}}{D^{{p_1}}_{{a_1}}}{D^{{i_2}{i_1}}_{{p_2}{p_3}}}} \nonumber \\
  & + {{{\frac{1}{2}}}{D^{{p_1}}_{{p_3}}}{D^{{i_1}{i_2}}_{{p_2}{a_1}}}} - {{D^{{i_2}}_{{a_1}}}{D^{{p_1}{i_1}}_{{p_2}{p_3}}}} + {{D^{{p_1}{i_1}}_{{p_2}{p_3}}}{\hat{E}^{{i_2}}_{{a_1}}}} \nonumber \\
  & - {{{\frac{1}{2}}}{D^{{p_1}{i_2}}_{{p_2}{p_3}}}{\hat{E}^{{i_1}}_{{a_1}}}} + {{{4}}{D^{{p_1}}_{{p_2}}}{D^{{i_1}}_{{p_3}}}{D^{{i_2}}_{{a_1}}}} - {{{2}}{D^{{p_1}}_{{p_2}}}{D^{{i_2}}_{{a_1}}}{\hat{E}^{{i_1}}_{{p_3}}}} \nonumber \\
  & - {{{\frac{1}{2}}}{D^{{i_2}}_{{p_2}}}{D^{{i_1}}_{{a_1}}}{\hat{E}^{{p_1}}_{{p_3}}}} 
 + {{D^{{p_1}}_{{p_3}}}{D^{{i_2}}_{{a_1}}}{\hat{E}^{{i_1}}_{{p_2}}}} - {{{2}}{D^{{i_1}}_{{p_3}}}{D^{{i_2}}_{{a_1}}}{\hat{E}^{{p_1}}_{{p_2}}}} \nonumber \\
  & + {{D^{{p_1}}_{{p_2}}}{D^{{i_2}}_{{p_3}}}{\hat{E}^{{i_1}}_{{a_1}}}} - {{{\frac{1}{2}}}{D^{{i_2}}_{{p_2}}}{D^{{p_1}}_{{p_3}}}{\hat{E}^{{i_1}}_{{a_1}}}} - {{{\frac{1}{2}}}{D^{{i_1}}_{{p_2}}}{D^{{p_1}}_{{a_1}}}{\hat{E}^{{i_2}}_{{p_3}}}} \nonumber \\
  & - {{{2}}{D^{{i_2}}_{{p_2}}}{D^{{i_1}}_{{p_3}}}{D^{{p_1}}_{{a_1}}}} - {{{\frac{1}{2}}}{D^{{p_1}}_{{p_3}}}{D^{{i_1}}_{{a_1}}}{\hat{E}^{{i_2}}_{{p_2}}}} + {{D^{{i_2}}_{{p_3}}}{D^{{i_1}}_{{a_1}}}{\hat{E}^{{p_1}}_{{p_2}}}} \nonumber \\
  & + {{D^{{i_2}}_{{p_2}}}{D^{{i_1}}_{{p_3}}}{\hat{E}^{{p_1}}_{{a_1}}}} + {{D^{{p_1}}_{{p_2}}}{D^{{i_1}}_{{a_1}}}{\hat{E}^{{i_2}}_{{p_3}}}} + {{D^{{i_1}}_{{p_3}}}{D^{{p_1}}_{{a_1}}}{\hat{E}^{{i_2}}_{{p_2}}}} \nonumber \\
  & - {{{2}}{D^{{p_1}}_{{p_2}}}{D^{{i_1}}_{{p_3}}}{\hat{E}^{{i_2}}_{{a_1}}}} - {{{\frac{1}{2}}}{D^{{i_1}}_{{p_2}}}{D^{{i_2}}_{{p_3}}}{\hat{E}^{{p_1}}_{{a_1}}}} + {{D^{{i_1}}_{{p_2}}}{D^{{i_2}}_{{p_3}}}{D^{{p_1}}_{{a_1}}}} \nonumber \\
  & - {{{2}}{D^{{p_1}}_{{p_2}}}{D^{{i_2}}_{{p_3}}}{D^{{i_1}}_{{a_1}}}} + {{D^{{i_2}}_{{p_2}}}{D^{{p_1}}_{{p_3}}}{D^{{i_1}}_{{a_1}}}} + {{D^{{i_1}}_{{p_2}}}{D^{{p_1}}_{{p_3}}}{\hat{E}^{{i_2}}_{{a_1}}}} \nonumber \\
  & - {{{\frac{1}{2}}}{D^{{i_2}}_{{p_3}}}{D^{{p_1}}_{{a_1}}}{\hat{E}^{{i_1}}_{{p_2}}}} - {{{2}}{D^{{i_1}}_{{p_2}}}{D^{{p_1}}_{{p_3}}}{D^{{i_2}}_{{a_1}}}} + {{D^{{i_1}}_{{p_2}}}{D^{{i_2}}_{{a_1}}}{\hat{E}^{{p_1}}_{{p_3}}}} \nonumber \\
  & + {{D^{{i_2}}_{{p_2}}}{D^{{p_1}}_{{a_1}}}{\hat{E}^{{i_1}}_{{p_3}}}} \bigg)
  \end{align}
The other alternative, ({\bf a+b2}), is obtained by invoking the cumulant decomposition first:
\begin{align}
\hat{E}_{p^{\prime}_{3}a_{1}p^{\prime}_{2}}^{i_1 i_2 p^{\prime}_1} \overset{\rm cm}{\approx}&  \bigg( - {{{\frac{1}{2}}}{D^{{i_2}}_{{p^{\prime}_3}}}{\hat{E}^{{p^{\prime}_1}{i_1}}_{{p^{\prime}_2}{a_1}}}} + {{{\frac{1}{2}}}{D^{{i_1}}_{{a_1}}}{D^{{p^{\prime}_1}{i_2}}_{{p^{\prime}_2}{p^{\prime}_3}}}} \nonumber \\
  & + {{D^{{p^{\prime}_1}{i_2}}_{{p^{\prime}_2}{a_1}}}{\hat{E}^{{i_1}}_{{p^{\prime}_3}}}} - {{{\frac{1}{2}}}{D^{{i_1}}_{{p^{\prime}_2}}}{\hat{E}^{{p^{\prime}_1}{i_2}}_{{p^{\prime}_3}{a_1}}}} + {{D^{{i_1}}_{{p^{\prime}_3}}}{\hat{E}^{{p^{\prime}_1}{i_2}}_{{p^{\prime}_2}{a_1}}}} \nonumber \\
  & - {{D^{{p^{\prime}_1}}_{{p^{\prime}_2}}}{D^{{i_1}{i_2}}_{{p^{\prime}_3}{a_1}}}} - {{{\frac{1}{2}}}{D^{{i_1}{p^{\prime}_1}}_{{p^{\prime}_3}{a_1}}}{\hat{E}^{{i_2}}_{{p^{\prime}_2}}}} - {{D^{{i_1}}_{{p^{\prime}_3}}}{D^{{p^{\prime}_1}{i_2}}_{{p^{\prime}_2}{a_1}}}} \nonumber \\
  & + {{{\frac{1}{2}}}{D^{{i_1}}_{{p^{\prime}_2}}}{D^{{p^{\prime}_1}{i_2}}_{{p^{\prime}_3}{a_1}}}} - {{{\frac{1}{2}}}{D^{{i_2}}_{{p^{\prime}_2}}}{\hat{E}^{{i_1}{p^{\prime}_1}}_{{p^{\prime}_3}{a_1}}}} - {{{\frac{1}{2}}}{D^{{p^{\prime}_1}{i_2}}_{{p^{\prime}_3}{a_1}}}{\hat{E}^{{i_1}}_{{p^{\prime}_2}}}} \nonumber \\
  & + {{{\frac{1}{2}}}{D^{{i_2}}_{{p^{\prime}_2}}}{D^{{i_1}{p^{\prime}_1}}_{{p^{\prime}_3}{a_1}}}} - {{{\frac{1}{2}}}{D^{{p^{\prime}_1}{i_1}}_{{p^{\prime}_2}{a_1}}}{\hat{E}^{{i_2}}_{{p^{\prime}_3}}}} - {{{\frac{1}{2}}}{D^{{i_2}{i_1}}_{{p^{\prime}_2}{p^{\prime}_3}}}{\hat{E}^{{p^{\prime}_1}}_{{a_1}}}} \nonumber \\
  & - {{{\frac{1}{2}}}{D^{{i_1}{i_2}}_{{p^{\prime}_2}{a_1}}}{\hat{E}^{{p^{\prime}_1}}_{{p^{\prime}_3}}}} - {{{\frac{1}{2}}}{D^{{p^{\prime}_1}}_{{a_1}}}{\hat{E}^{{i_2}{i_1}}_{{p^{\prime}_2}{p^{\prime}_3}}}} - {{{\frac{1}{2}}}{D^{{p^{\prime}_1}}_{{p^{\prime}_3}}}{\hat{E}^{{i_1}{i_2}}_{{p^{\prime}_2}{a_1}}}} \nonumber \\
  & + {{D^{{i_2}}_{{a_1}}}{\hat{E}^{{p^{\prime}_1}{i_1}}_{{p^{\prime}_2}{p^{\prime}_3}}}} + {{D^{{i_1}{i_2}}_{{p^{\prime}_3}{a_1}}}{\hat{E}^{{p^{\prime}_1}}_{{p^{\prime}_2}}}} - {{{\frac{1}{2}}}{D^{{i_1}}_{{a_1}}}{\hat{E}^{{p^{\prime}_1}{i_2}}_{{p^{\prime}_2}{p^{\prime}_3}}}} \nonumber \\
  & + {{{\frac{1}{2}}}{D^{{i_2}}_{{p^{\prime}_3}}}{D^{{p^{\prime}_1}{i_1}}_{{p^{\prime}_2}{a_1}}}} + {{D^{{p^{\prime}_1}}_{{p^{\prime}_2}}}{\hat{E}^{{i_1}{i_2}}_{{p^{\prime}_3}{a_1}}}} + {{{\frac{1}{2}}}{D^{{p^{\prime}_1}}_{{a_1}}}{D^{{i_2}{i_1}}_{{p^{\prime}_2}{p^{\prime}_3}}}} \nonumber \\
  & + {{{\frac{1}{2}}}{D^{{p^{\prime}_1}}_{{p^{\prime}_3}}}{D^{{i_1}{i_2}}_{{p^{\prime}_2}{a_1}}}} - {{D^{{i_2}}_{{a_1}}}{D^{{p^{\prime}_1}{i_1}}_{{p^{\prime}_2}{p^{\prime}_3}}}} + {{D^{{p^{\prime}_1}{i_1}}_{{p^{\prime}_2}{p^{\prime}_3}}}{\hat{E}^{{i_2}}_{{a_1}}}} \nonumber \\
  & - {{{\frac{1}{2}}}{D^{{p^{\prime}_1}{i_2}}_{{p^{\prime}_2}{p^{\prime}_3}}}{\hat{E}^{{i_1}}_{{a_1}}}} + {{{4}}{D^{{p^{\prime}_1}}_{{p^{\prime}_2}}}{D^{{i_1}}_{{p^{\prime}_3}}}{D^{{i_2}}_{{a_1}}}} - {{{2}}{D^{{p^{\prime}_1}}_{{p^{\prime}_2}}}{D^{{i_2}}_{{a_1}}}{\hat{E}^{{i_1}}_{{p^{\prime}_3}}}} \nonumber \\
  & - {{{\frac{1}{2}}}{D^{{i_2}}_{{p^{\prime}_2}}}{D^{{i_1}}_{{a_1}}}{\hat{E}^{{p^{\prime}_1}}_{{p^{\prime}_3}}}}  + {{D^{{p^{\prime}_1}}_{{p^{\prime}_3}}}{D^{{i_2}}_{{a_1}}}{\hat{E}^{{i_1}}_{{p^{\prime}_2}}}} - {{{2}}{D^{{i_1}}_{{p^{\prime}_3}}}{D^{{i_2}}_{{a_1}}}{\hat{E}^{{p^{\prime}_1}}_{{p^{\prime}_2}}}} \nonumber \\
  & + {{D^{{p^{\prime}_1}}_{{p^{\prime}_2}}}{D^{{i_2}}_{{p^{\prime}_3}}}{\hat{E}^{{i_1}}_{{a_1}}}} - {{{\frac{1}{2}}}{D^{{i_2}}_{{p^{\prime}_2}}}{D^{{p^{\prime}_1}}_{{p^{\prime}_3}}}{\hat{E}^{{i_1}}_{{a_1}}}} - {{{\frac{1}{2}}}{D^{{i_1}}_{{p^{\prime}_2}}}{D^{{p^{\prime}_1}}_{{a_1}}}{\hat{E}^{{i_2}}_{{p^{\prime}_3}}}} \nonumber \\
  & - {{{2}}{D^{{i_2}}_{{p^{\prime}_2}}}{D^{{i_1}}_{{p^{\prime}_3}}}{D^{{p^{\prime}_1}}_{{a_1}}}} - {{{\frac{1}{2}}}{D^{{p^{\prime}_1}}_{{p^{\prime}_3}}}{D^{{i_1}}_{{a_1}}}{\hat{E}^{{i_2}}_{{p^{\prime}_2}}}} + {{D^{{i_2}}_{{p^{\prime}_3}}}{D^{{i_1}}_{{a_1}}}{\hat{E}^{{p^{\prime}_1}}_{{p^{\prime}_2}}}} \nonumber \\
  & + {{D^{{i_2}}_{{p^{\prime}_2}}}{D^{{i_1}}_{{p^{\prime}_3}}}{\hat{E}^{{p^{\prime}_1}}_{{a_1}}}} + {{D^{{p^{\prime}_1}}_{{p^{\prime}_2}}}{D^{{i_1}}_{{a_1}}}{\hat{E}^{{i_2}}_{{p^{\prime}_3}}}} + {{D^{{i_1}}_{{p^{\prime}_3}}}{D^{{p^{\prime}_1}}_{{a_1}}}{\hat{E}^{{i_2}}_{{p^{\prime}_2}}}} \nonumber \\
  & - {{{2}}{D^{{p^{\prime}_1}}_{{p^{\prime}_2}}}{D^{{i_1}}_{{p^{\prime}_3}}}{\hat{E}^{{i_2}}_{{a_1}}}} - {{{\frac{1}{2}}}{D^{{i_1}}_{{p^{\prime}_2}}}{D^{{i_2}}_{{p^{\prime}_3}}}{\hat{E}^{{p^{\prime}_1}}_{{a_1}}}} + {{D^{{i_1}}_{{p^{\prime}_2}}}{D^{{i_2}}_{{p^{\prime}_3}}}{D^{{p^{\prime}_1}}_{{a_1}}}} \nonumber \\
  & - {{{2}}{D^{{p^{\prime}_1}}_{{p^{\prime}_2}}}{D^{{i_2}}_{{p^{\prime}_3}}}{D^{{i_1}}_{{a_1}}}} + {{D^{{i_2}}_{{p^{\prime}_2}}}{D^{{p^{\prime}_1}}_{{p^{\prime}_3}}}{D^{{i_1}}_{{a_1}}}} + {{D^{{i_1}}_{{p^{\prime}_2}}}{D^{{p^{\prime}_1}}_{{p^{\prime}_3}}}{\hat{E}^{{i_2}}_{{a_1}}}} \nonumber \\
  & - {{{\frac{1}{2}}}{D^{{i_2}}_{{p^{\prime}_3}}}{D^{{p^{\prime}_1}}_{{a_1}}}{\hat{E}^{{i_1}}_{{p^{\prime}_2}}}} - {{{2}}{D^{{i_1}}_{{p^{\prime}_2}}}{D^{{p^{\prime}_1}}_{{p^{\prime}_3}}}{D^{{i_2}}_{{a_1}}}} + {{D^{{i_1}}_{{p^{\prime}_2}}}{D^{{i_2}}_{{a_1}}}{\hat{E}^{{p^{\prime}_1}}_{{p^{\prime}_3}}}} \nonumber \\
  & + {{D^{{i_2}}_{{p^{\prime}_2}}}{D^{{p^{\prime}_1}}_{{a_1}}}{\hat{E}^{{i_1}}_{{p^{\prime}_3}}}}\bigg)
  \end{align}
The subsequent exclusion of the core orbitals replaces core-including OBS orbital indices on the $\hat{E}$ operators only, but not on the RDMs:
 \begin{align}
 \label{eq:E3-b2}
\hat{E}_{p^{\prime}_{3}a_{1}p^{\prime}_{2}}^{i_1 i_2 p^{\prime}_1} \overset{\rm \mathbf{b2}=cm+fzc}{\approx}&  \bigg( - {{{\frac{1}{2}}}{D^{{i_2}}_{{p^{\prime}_3}}}{\hat{E}^{{p_1}{i_1}}_{{p_2}{a_1}}}} + {{{\frac{1}{2}}}{D^{{i_1}}_{{a_1}}}{D^{{p^{\prime}_1}{i_2}}_{{p^{\prime}_2}{p^{\prime}_3}}}} \nonumber \\
  & + {{D^{{p^{\prime}_1}{i_2}}_{{p^{\prime}_2}{a_1}}}{\hat{E}^{{i_1}}_{{p_3}}}} - {{{\frac{1}{2}}}{D^{{i_1}}_{{p^{\prime}_2}}}{\hat{E}^{{p_1}{i_2}}_{{p_3}{a_1}}}} + {{D^{{i_1}}_{{p^{\prime}_3}}}{\hat{E}^{{p_1}{i_2}}_{{p_2}{a_1}}}} \nonumber \\
  & - {{D^{{p^{\prime}_1}}_{{p^{\prime}_2}}}{D^{{i_1}{i_2}}_{{p^{\prime}_3}{a_1}}}} - {{{\frac{1}{2}}}{D^{{i_1}{p^{\prime}_1}}_{{p^{\prime}_3}{a_1}}}{\hat{E}^{{i_2}}_{{p_2}}}} - {{D^{{i_1}}_{{p^{\prime}_3}}}{D^{{p^{\prime}_1}{i_2}}_{{p^{\prime}_2}{a_1}}}} \nonumber \\
  & + {{{\frac{1}{2}}}{D^{{i_1}}_{{p^{\prime}_2}}}{D^{{p^{\prime}_1}{i_2}}_{{p^{\prime}_3}{a_1}}}} - {{{\frac{1}{2}}}{D^{{i_2}}_{{p^{\prime}_2}}}{\hat{E}^{{i_1}{p_1}}_{{p_3}{a_1}}}} - {{{\frac{1}{2}}}{D^{{p^{\prime}_1}{i_2}}_{{p^{\prime}_3}{a_1}}}{\hat{E}^{{i_1}}_{{p_2}}}} \nonumber \\
  & + {{{\frac{1}{2}}}{D^{{i_2}}_{{p^{\prime}_2}}}{D^{{i_1}{p^{\prime}_1}}_{{p^{\prime}_3}{a_1}}}} - {{{\frac{1}{2}}}{D^{{p^{\prime}_1}{i_1}}_{{p^{\prime}_2}{a_1}}}{\hat{E}^{{i_2}}_{{p_3}}}} - {{{\frac{1}{2}}}{D^{{i_2}{i_1}}_{{p^{\prime}_2}{p^{\prime}_3}}}{\hat{E}^{{p_1}}_{{a_1}}}} \nonumber \\
  & - {{{\frac{1}{2}}}{D^{{i_1}{i_2}}_{{p^{\prime}_2}{a_1}}}{\hat{E}^{{p_1}}_{{p_3}}}} - {{{\frac{1}{2}}}{D^{{p^{\prime}_1}}_{{a_1}}}{\hat{E}^{{i_2}{i_1}}_{{p_2}{p_3}}}} - {{{\frac{1}{2}}}{D^{{p^{\prime}_1}}_{{p^{\prime}_3}}}{\hat{E}^{{i_1}{i_2}}_{{p_2}{a_1}}}} \nonumber \\
  & + {{D^{{i_2}}_{{a_1}}}{\hat{E}^{{p_1}{i_1}}_{{p_2}{p_3}}}} + {{D^{{i_1}{i_2}}_{{p^{\prime}_3}{a_1}}}{\hat{E}^{{p_1}}_{{p_2}}}} - {{{\frac{1}{2}}}{D^{{i_1}}_{{a_1}}}{\hat{E}^{{p_1}{i_2}}_{{p_2}{p_3}}}} \nonumber \\
  & + {{{\frac{1}{2}}}{D^{{i_2}}_{{p^{\prime}_3}}}{D^{{p^{\prime}_1}{i_1}}_{{p^{\prime}_2}{a_1}}}} + {{D^{{p^{\prime}_1}}_{{p^{\prime}_2}}}{\hat{E}^{{i_1}{i_2}}_{{p_3}{a_1}}}} + {{{\frac{1}{2}}}{D^{{p^{\prime}_1}}_{{a_1}}}{D^{{i_2}{i_1}}_{{p^{\prime}_2}{p^{\prime}_3}}}} \nonumber \\
  & + {{{\frac{1}{2}}}{D^{{p^{\prime}_1}}_{{p^{\prime}_3}}}{D^{{i_1}{i_2}}_{{p^{\prime}_2}{a_1}}}} - {{D^{{i_2}}_{{a_1}}}{D^{{p^{\prime}_1}{i_1}}_{{p^{\prime}_2}{p^{\prime}_3}}}} + {{D^{{p^{\prime}_1}{i_1}}_{{p^{\prime}_2}{p^{\prime}_3}}}{\hat{E}^{{i_2}}_{{a_1}}}} \nonumber \\
  & - {{{\frac{1}{2}}}{D^{{p^{\prime}_1}{i_2}}_{{p^{\prime}_2}{p^{\prime}_3}}}{\hat{E}^{{i_1}}_{{a_1}}}} + {{{4}}{D^{{p^{\prime}_1}}_{{p^{\prime}_2}}}{D^{{i_1}}_{{p^{\prime}_3}}}{D^{{i_2}}_{{a_1}}}} - {{{2}}{D^{{p^{\prime}_1}}_{{p^{\prime}_2}}}{D^{{i_2}}_{{a_1}}}{\hat{E}^{{i_1}}_{{p_3}}}} \nonumber \\
  & - {{{\frac{1}{2}}}{D^{{i_2}}_{{p^{\prime}_2}}}{D^{{i_1}}_{{a_1}}}{\hat{E}^{{p_1}}_{{p_3}}}}  + {{D^{{p^{\prime}_1}}_{{p^{\prime}_3}}}{D^{{i_2}}_{{a_1}}}{\hat{E}^{{i_1}}_{{p_2}}}} - {{{2}}{D^{{i_1}}_{{p^{\prime}_3}}}{D^{{i_2}}_{{a_1}}}{\hat{E}^{{p_1}}_{{p_2}}}} \nonumber \\
  & + {{D^{{p^{\prime}_1}}_{{p^{\prime}_2}}}{D^{{i_2}}_{{p^{\prime}_3}}}{\hat{E}^{{i_1}}_{{a_1}}}} - {{{\frac{1}{2}}}{D^{{i_2}}_{{p^{\prime}_2}}}{D^{{p^{\prime}_1}}_{{p^{\prime}_3}}}{\hat{E}^{{i_1}}_{{a_1}}}} - {{{\frac{1}{2}}}{D^{{i_1}}_{{p^{\prime}_2}}}{D^{{p^{\prime}_1}}_{{a_1}}}{\hat{E}^{{i_2}}_{{p_3}}}} \nonumber \\
  & - {{{2}}{D^{{i_2}}_{{p^{\prime}_2}}}{D^{{i_1}}_{{p^{\prime}_3}}}{D^{{p^{\prime}_1}}_{{a_1}}}} - {{{\frac{1}{2}}}{D^{{p^{\prime}_1}}_{{p^{\prime}_3}}}{D^{{i_1}}_{{a_1}}}{\hat{E}^{{i_2}}_{{p_2}}}} + {{D^{{i_2}}_{{p^{\prime}_3}}}{D^{{i_1}}_{{a_1}}}{\hat{E}^{{p_1}}_{{p_2}}}} \nonumber \\
  & + {{D^{{i_2}}_{{p^{\prime}_2}}}{D^{{i_1}}_{{p^{\prime}_3}}}{\hat{E}^{{p_1}}_{{a_1}}}} + {{D^{{p^{\prime}_1}}_{{p^{\prime}_2}}}{D^{{i_1}}_{{a_1}}}{\hat{E}^{{i_2}}_{{p_3}}}} + {{D^{{i_1}}_{{p^{\prime}_3}}}{D^{{p^{\prime}_1}}_{{a_1}}}{\hat{E}^{{i_2}}_{{p_2}}}} \nonumber \\
  & - {{{2}}{D^{{p^{\prime}_1}}_{{p^{\prime}_2}}}{D^{{i_1}}_{{p^{\prime}_3}}}{\hat{E}^{{i_2}}_{{a_1}}}} - {{{\frac{1}{2}}}{D^{{i_1}}_{{p^{\prime}_2}}}{D^{{i_2}}_{{p^{\prime}_3}}}{\hat{E}^{{p_1}}_{{a_1}}}} + {{D^{{i_1}}_{{p^{\prime}_2}}}{D^{{i_2}}_{{p^{\prime}_3}}}{D^{{p^{\prime}_1}}_{{a_1}}}} \nonumber \\
  & - {{{2}}{D^{{p^{\prime}_1}}_{{p^{\prime}_2}}}{D^{{i_2}}_{{p^{\prime}_3}}}{D^{{i_1}}_{{a_1}}}} + {{D^{{i_2}}_{{p^{\prime}_2}}}{D^{{p^{\prime}_1}}_{{p^{\prime}_3}}}{D^{{i_1}}_{{a_1}}}} + {{D^{{i_1}}_{{p^{\prime}_2}}}{D^{{p^{\prime}_1}}_{{p^{\prime}_3}}}{\hat{E}^{{i_2}}_{{a_1}}}} \nonumber \\
  & - {{{\frac{1}{2}}}{D^{{i_2}}_{{p^{\prime}_3}}}{D^{{p^{\prime}_1}}_{{a_1}}}{\hat{E}^{{i_1}}_{{p_2}}}} - {{{2}}{D^{{i_1}}_{{p^{\prime}_2}}}{D^{{p^{\prime}_1}}_{{p^{\prime}_3}}}{D^{{i_2}}_{{a_1}}}} + {{D^{{i_1}}_{{p^{\prime}_2}}}{D^{{i_2}}_{{a_1}}}{\hat{E}^{{p_1}}_{{p_3}}}} \nonumber \\
  & + {{D^{{i_2}}_{{p^{\prime}_2}}}{D^{{p^{\prime}_1}}_{{a_1}}}{\hat{E}^{{i_1}}_{{p_3}}}}\bigg)
  \end{align}
Note the appearance of the core-including OBS indices in the ({\bf b2}) approximant for $\hat{E}_{p^{\prime}_{3}a_{1}p^{\prime}_{2}}^{i_1 i_2 p^{\prime}_1}$ (\cref{eq:E3-b2}) but not in the in the corresponding ({\bf b1}) expression (\cref{eq:E3-b1}).

\bibliography{vrgrefs}